\documentclass{aastex63}

\usepackage{lineno}
\usepackage{ulem}

 \def\mso{\,\mathrm{M}_\odot}

 \def\simle{\mathrel{\hbox{\rlap{\hbox{\lower4pt\hbox{$\sim$}}}\hbox{$<$}}}}
 \def\simgr{\mathrel{\hbox{\rlap{\hbox{\lower4pt\hbox{$\sim$}}}\hbox{$>$}}}}
\renewcommand{\thefigure}{\arabic{figure}}

\received{2019 November 18}
\revised{2019 December 10}
\accepted{2019 December 11}
\submitjournal{ApJL}

\shorttitle{Effects of close binary evolution on the main-sequence morphology of young star clusters}
\shortauthors{Chen Wang et al.}

\begin{document}

\title{Effects of close binary evolution on the main-sequence morphology of young star clusters}

\correspondingauthor{Chen Wang}
\email{cwang@astro.uni-bonn.de}

\author[0000-0002-0716-3801]{Chen Wang}
\affiliation{Argelander-Institut f\"ur Astronomie, Universit\"at Bonn, Auf dem H\"ugel 71, 53121 Bonn, Germany}

\author{Norbert Langer}
\affiliation{Argelander-Institut f\"ur Astronomie, Universit\"at Bonn, Auf dem H\"ugel 71, 53121 Bonn, Germany}

\author[0000-0002-2715-7484]{Abel Schootemeijer}
\affiliation{Argelander-Institut f\"ur Astronomie, Universit\"at Bonn, Auf dem H\"ugel 71, 53121 Bonn, Germany}

\author[0000-0003-0521-473X]{Norberto Castro}
\affiliation{AIP Potsdam, An der Sternwarte 16, 14482 Potsdam, Germany}

\author{Sylvia Adscheid}
\affiliation{Argelander-Institut f\"ur Astronomie, Universit\"at Bonn, Auf dem H\"ugel 71, 53121 Bonn, Germany}

\author[0000-0002-0338-8181]{Pablo Marchant}
\affiliation{Department of Physics and Astronomy, Northwestern University, 2145 Sheridan Road, Evanston, IL 60208, USA}

\author{Ben Hastings}
\affiliation{Argelander-Institut f\"ur Astronomie, Universit\"at Bonn, Auf dem H\"ugel 71, 53121 Bonn, Germany}

\begin{abstract}

Star clusters are the  building blocks of galaxies. They are composed of stars of
nearly equal age and chemical composition, allowing us to use them as chronometers
and as testbeds for gauging stellar evolution. It has become clear recently that
massive stars are formed preferentially in close binaries, in which mass transfer 
will drastically change the evolution of the stars. This is expected to leave a significant
imprint in the distribution of cluster stars in the Hertzsprung-Russell diagram. Our results,
based on a dense model grid of more than 50,000 detailed binary-evolution calculations,
indeed show several distinct, coeval  main-sequence (MS) components, most notably an extended  MS
turnoff region, and a group of near-critical rotating stars that is spread
over a large luminosity range on the red side of the classical MS.
We comprehensively demonstrate the time evolution of the features in an animation,
and we derive analytic expressions to describe these features. We find quantitative agreement
with results based on recent photometric and spectroscopic observations. We conclude that while other
factors may also be at play, binary evolution has a major impact on the MS
morphology of young star clusters.
\end{abstract}

\section{Introduction} \label{sec:intro}
The stellar main sequence (MS) in the Hertzsprung-Russell (HR) diagram,
which was discovered \citep{Russell1914} before it was known that stars are powered by nuclear fusion \citep{1929ZPhy...54..656A},
is the backbone of our knowledge of stars. Until today, the analysis of
star clusters has given essential clues for understanding the internal structure and evolution of stars.
Research in recent years has shown that
the MSs of star clusters are not simple one-dimensional structures; rather, they contain distinct features that still need to be deciphered.

In studies of star clusters, frequent key assumptions are that the stars in a cluster are coeval, 
have the same initial chemical composition, are not affected by rotation, and are single stars. 
All four assumptions are currently being challenged by modern, high-precision observations. 
Recently, photometric studies using the Hubble Space Telescope
revealed that the MS of young and intermediate age star clusters (14-600\,Myr) in the Magellanic Clouds
and in the Galaxy is split into two distinct MSs, as well as an extended MS turnoff 
region \citep[][ and references therein]{2017NatAs...1E.186D,Milone2018,2019ApJ...876...65L}. 
Furthermore, the young clusters show distinct groups of emission-line stars, extending more than two magnitudes 
below the turnoff \citep{Milone2018}, most of which are spectroscopically identified as Be stars
\citep{bodensteiner2019young}. 


In this situation, it appears worthwhile to investigate binary-evolution effects.   
The observed binary incidence and binary properties of massive stars imply that an isolated life of 
a massive star is the exception, the rule being that its evolution will be strongly affected by a binary
companion \citep{2012Sci...337..444S}. Simplified binary population synthesis calculations
in which binary evolution is approximated based on single-star models, have shown that binary evolution can produce 
a rapidly rotating sub-population \citep{2013ApJ...764..166D}, and that it can account for the so-called blue stragglers, i.e.,
cluster stars above the apparent cluster turnoff, in Galactic open clusters with ages of 
up to $\sim 1\,$Gyr \citep{2011ApJ...731L..37Y,2015ApJ...805...20S}.  
However, such so called rapid binary-evolution models are not able to predict reliable effective temperatures of post-interaction stars.
To understand the MS morphology of young star clusters, it is therefore necessary to
investigate the effects of close binary evolution based on dense grids of detailed binary-evolution models \citep{1998A&A...334...21V}. 

Here, we provide such grids, which
include differential rotation, rotationally induced internal mixing, and magnetic angular momentum transport
as used by \citet{2017A&A...604A..55M} and use them to investigate the effects of 
close binary evolution on the MS morphology of young star clusters. 
In Sect.\,2 we describe the adopted physics to compute binary models, and we present our main results in Sect.\,3.
We provide a comparison with single-star models in Sect.\,4, and with observed star clusters in Sect.\,5.
In Sect.\,6, we give our concluding remarks.

\section{Method and assumptions}\label{sec:method}
We use the detailed one-dimensional stellar evolution code MESA \citep{Paxton2011,Paxton2013,Paxton2015}, version 8845, 
to compute our binary-evolution models. The stellar models contain physics assumptions that are 
identical to the rotating single-star models of \citet{2011A&A...530A.115B}, and include 
differential rotation, rotationally induced internal mixing, magnetic angular momentum transport,
stellar wind mass loss, and non-equilibrium CNO nucleosynthesis. Our assumptions on binary
physics are as those described by \citet{2017A&A...604A..55M}. 
Below, we emphasize the most relevant physical assumptions for convenience.

We use the standard mixing-length theory to model convection with a mixing-length parameter of $\alpha = l/H_{P} = 1.5$,
where $H_{P}$ denotes the local pressure scale height, which allows for inflated envelopes in models near their
Eddington limit \citep{2015A&A...580A..20S,2017A&A...597A..71S}.
To determine the boundaries of convective zones, we adopt the Ledoux criterion,
where we include convective core overshooting as a step function with $\alpha_{\mathrm{OV}}=0.335$ \citep{2011A&A...530A.115B}.
We include semiconvective mixing in superadiabatic layers if they are stable according to the Ledoux criterion
using $\alpha_{\mathrm{sc}}=1$ \citep{1983A&A...126..207L}, and thermohaline mixing as used by \citet{2010A&A...521A...9C}
with $\alpha_{\mathrm{th}}=1$. Rotationally induced mixing is
modeled as a diffusive process \citep{2000ApJ...528..368H}. We take into account the effects of
dynamical and secular shear instabilities, the Goldreich-Schubert-Fricke instability, and Eddington-Sweet circulations.
The efficiency parameter of rotational mixing is $f_{c}=1/30$ as proposed by \citet{1992A&A...253..173C}.
We also include the Tayler-Spruit dynamo for the transport of angular momentum \citep{Spruit2002,Heger2005}.
 
The detailed structure of both binary components is computed simultaneously with the orbital evolution.
We assume circular orbits and adopt initial spins such that the rotation period of both stars is synchronized to the orbital period of the binary on the zero-age MS. During the evolution, the effects of tidally induced spin-orbit coupling \citep{2008A&A...484..831D} are included. 
We compute mass and angular momentum transfer arising from Roche-lobe overflow.
We assume the specific angular momentum accreted by the secondary star depends
on whether the accretion is ballistic or occurs via a Keplerian disk, and
restrict the mass accretion of the mass gainer when it has reached near-critical rotation \citep{2005A&A...435.1013P}.
We do so by adopting a rotational enhancement of the stellar mass-loss rate, which prevents models from exceeding critical rotation 
\citep{1998A&A...329..551L,2012ARA&A..50..107L,Paxton2015}. 
When the energy contained in the combined luminosity of both stars is insufficient to drive the mass out of the binary at a rate equal to the mass loss rate, we assume that both stars are engulfed in the excess material, with a binary merger ensuing.

For contact phases, we employ the scheme described by \citet{Pablo2016} to model the mass-transfer  phase.
We assume both stars to merge when they both fill their Roche volumes and mass outflow through the second Lagrangian point is obtained. 
When a binary merger is assumed to happen for a given binary model while both stars still burn hydrogen in their cores,
we calculate its further evolution by adopting a single-star model with the appropriate mass and age. We assume that the
internal structure fully rejuvenates, and adopt an initial central hydrogen abundance of the merger product according
to \citet{2016MNRAS.457.2355S}. Mergers are treated as non-rotating stars, following \cite{2019Natur.574..211S}.

The evolution of the binary models is started by considering both stars on the zero-age MS. 
We then follow the evolution of both components, as long as one of the two stars is still on the MS. 
We compute our models until core carbon exhaustion. If at that stage the core of a model 
exceeds the Chandrasekhar mass, we assume that it produces a supernova explosion, and compute the continued 
evolution of its core hydrogen burning companion in isolation.   

We adopt a metallicity of $Z_{\mathrm{SMC}}=0.002179$ appropriate for young stars in the
SMC, a helium abundance of $Y=0.25184$, and a distribution of heavy elements as used by \citet{2011A&A...530A.115B}.
Our model grid covers initial primary star masses between 5$\mso$ and 100$\mso$,
initial mass ratios of 0.95-0.3, and initial orbital periods of 1\,day-8.6\,yr.
We cover the initial parameter space with more than 50,000 binary-evolution models,
using 26 different initial primary masses and 140 initial period
values, both distributed evenly in log-space, and 14 evenly distributed initial mass ratios.

We compute a second suite of binary models in order to predict the distribution of stars in the HR diagram of a star cluster 
with a total mass of $10^5\,\mso$ in stars between $100\,\mso$ and $0.8\mso$.  
This leads to 2078 binary systems with primary masses larger than $5\,\mso$.
We use a Monte Carlo simulation to generate initial binary systems adopting a Salpeter initial mass function (IMF) \citep{Salpeter1955},
a flat distribution of initial mass rations $q_{\rm i}$, and a flat distribution of $\log\, P_i$, within the
initial mass ratio and period ranges as described above.
We assume a binary fraction of one, i.e., we do not consider additional single stars in our cluster model.
We use the MESA code to evolve the generated 2078 binary systems in time.
In this way, the time dependence of the HRD distribution of the cluster stars can be
simulated without the need to interpolate between different binary-evolution models. An interpolation in time is still necessary,
which, due to the high time resolution of the MESA models, does not lead to noticeable errors.

\section{Binary-induced main-sequence features}\label{sec:Results}
\begin{figure}
\centering
 \includegraphics[width=0.7\linewidth]{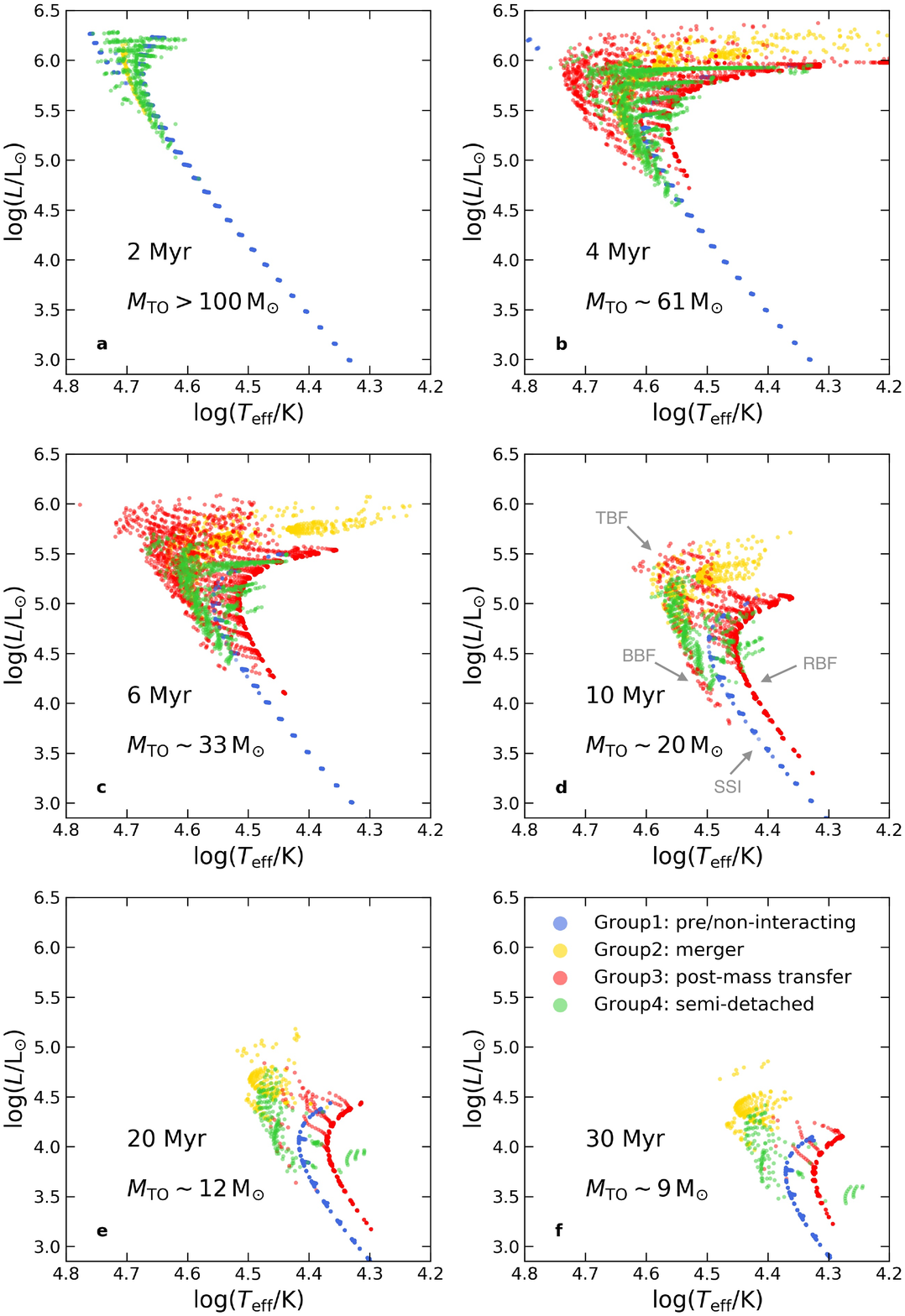}
\caption{Distribution of our binary models in the Hertzsprung-Russell diagram at six different times, as indicated.
Only the visually brighter component of each binary is plotted, for all models where at least 
one of the two stars still undergoes core hydrogen burning.
Dark blue dots correspond to models that have not yet interacted with a companion. Yellow dots indicate
the products of the merger of two core hydrogen burning stars. Red dots indicate stars that have undergone
rapid (thermal timescale) accretion of matter from a companion. Green dots represent stars undergoing
nuclear timescale accretion at the selected time. Below the age, we indicate the approximate MS turnoff mass
for the non-interacting stars. We indicate the binary produced MS features in Panel\,(d).
 \label{fig:HRD6}}
\end{figure}

Figure\,\ref{fig:HRD6} shows the locations of our models in the HR diagram
for selected ages. Each of the $\sim$50,000 detailed binary evolutionary sequences provides one dot
 in each figure, as long as at least one of the two stars still undergoes core hydrogen burning.
We only plot the visually brighter component of each system, as in most cases the fainter
one is either unevolved, lost its envelope, or terminated its evolution \citep{2014ApJ...782....7D}. 
Statistical probabilities due to the IMF or initial binary parameter distributions are not taken
into account, but the figure is meant to demonstrate which parts of the HR
diagram are covered by models in the different evolutionary branches.
In this figure, we distinguish models in four different evolutionary
branches, as indicated. 

The pre-interaction binaries (blue) in Fig.\,\ref{fig:HRD6} are located on a single line,
which represents the corresponding single-star isochrone (SSI).
As most of our pre-interaction stars rotate slowly by construction, the rotational
broadening of the SSI is small, even near the turnoff \citep[see Figure\,7 of][]{2011A&A...530A.115B}.
This ensures that all other features that are visible in Fig.\,\ref{fig:HRD6} are
induced by binary evolution, and not an effect of the initial distribution of rotational velocities. 

We distinguish three prominent features in the distribution of stars in Fig.\,\ref{fig:HRD6}. First, at all times, the turnoff 
region is extended beyond the SSI, mostly to higher luminosities and temperatures (we call this the turnoff binary-evolution feature, TBF). 
Second, for ages above $\sim 4$\,Myr, a distinct red MS appears ( red binary-evolution feature, RBF), 
whose lower part is well separated from the SSI. Third, at all times, MS stars to the blue side of the SSI appear. 
They form a separated blue MS for ages above $\sim 10\,$Myr and merge with the TBF at high luminosities 
(blue binary-evolution feature, BBF). We indicate these 
features in Panel\,(d) of Fig.\,\ref{fig:HRD6}. None of these three features has been described 
by detailed binary-evolution models before. 

\begin{figure}
\centering
\includegraphics[width=0.9\linewidth]{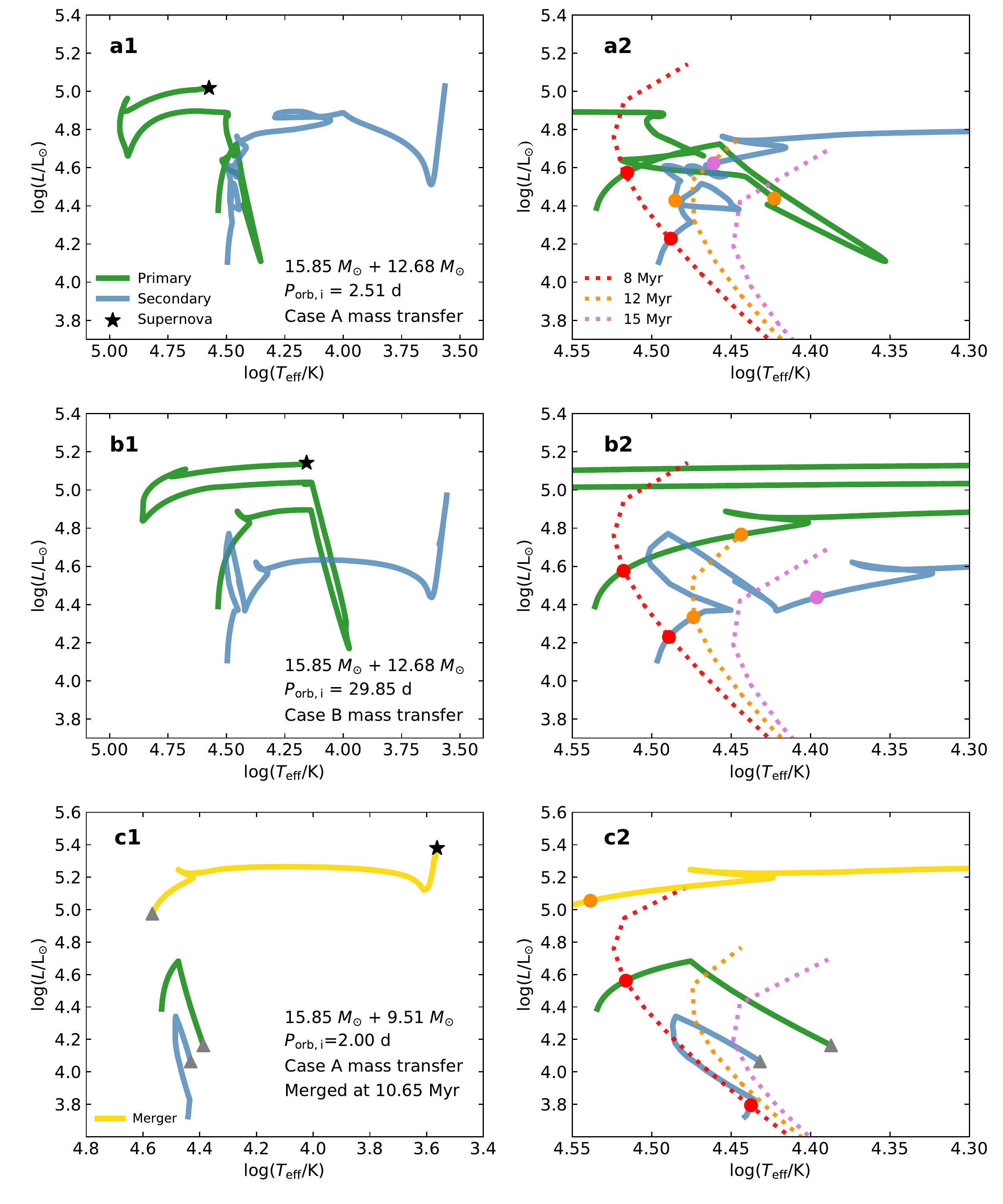}
 \caption{Evolution of both components of three example binary models in the HR diagram (left panels), 
with a zoom of the MS region (right panels).
All three models start with the same primary model (donor star, green lines) of 15.85$\mso$.
The initial mass of the secondary stars (mass gainer, blue lines) is 12.68$\mso$ in the top two systems,
and 9.51$\mso$ in the third. The dashed lines in the right panels represent single-star isochrones for 
8, 12, and 15\,Myr, and solid dots on the evolutionary tracks with the corresponding colors 
mark the positions of the binary components at these ages. The black star symbols mark the pre-supernova position
of the primary stars. The gray triangles in the bottom panels mark the merger event. 
The donor stars in the top and the bottom models start mass transfer during core hydrogen burning
(Case\,A mass transfer), and the donor in the middle model starts mass transfer after core hydrogen
exhaustion (Case\,B mass transfer).
\label{fig:Ex}}
\end{figure}

The evolution of exemplary binary models in the HR diagram in Fig.\,\ref{fig:Ex}
 helps reveal the origin of the MS features.
Binaries that undergo mass transfer while both stars fuse hydrogen in their cores (Case\,A mass transfer)
evolve into semi-detached systems. This occurs because Case\,A mass transfer comprises a nuclear timescale mass-transfer phase,
whereas otherwise mass transfer occurs on the much shorter thermal or even on the dynamical timescale.
As shown in Fig.\,\ref{fig:HRD6}, the semi-detached models do not give rise to a sharp feature in the HR diagram,
since their locations are quite spread out, and overlap strongly with those of the mergers and post-mass-transfer systems. 
These together produce the TBF. They are mostly located above and to the left of the line formed by the pre-interaction stars
because their mass ratios are inverted, since the former secondary stars have accreted substantial amounts
of matter in a preceding thermal timescale Case\,A mass transfer. If the secondary's mass has been
increased to values above the mass corresponding to the single-star turnoff,
it will be brighter than any of the pre-interaction stars at the given age. 
The internal process of rejuvenation, i.e., the mixing of fresh fuel
toward the centre due to an increase of the mass of the convective core, makes it look younger, i.e., leads to
a higher surface temperature than that of a comparable pre-interaction star \citep{2015ApJ...805...20S}.
After the mass donor in a Case\,A binary system ends its core hydrogen burning, its post-MS expansion
will give rise to another rapid, thermal timescale, mass-transfer phase, after which we consider the binary
a post-mass-transfer system.

Our Case\,A binaries produce a rich spectrum of observable features. The BBF, as described above,
is exclusively due to Case\,A mass transfer. 
Whereas models in the extended turnoff region are blue mainly because they are rejuvenated, 
the fainter blue MS stars of the BBF, which are most pronounced for ages in the range of 10-20\,Myr, 
have higher surface temperatures because their envelopes are significantly enriched in helium. 
They originate from the binaries with the shortest initial periods and
smallest initial mass ratios, which undergo contact evolution but avoid merging \citep{Pablo2016}.
Furthermore, we find some detached systems in between fast and slow Case\,A mass transfer,                             
which share their HR diagram positions with the semi-detached systems. At the highest masses,
strong stellar winds can lead to a widening of the orbit after the fast Case\,A mass transfer.
This may detach the mass donor from its Roche lobe for some time \citep{2005A&A...435.1013P}. At lower masses,
the orbital period of systems right after the fast Case\,A mass transfer can be short enough that 
tidally induced spin-orbit coupling leads to an expansion of the orbit, which again allows the primary stars 
to detach from their Roche lobe for about 0.1-0.2\,Myr before the slow Case\,A mass transfer begins.
We also find some semi-detached binary models in which the mass donor is still the brighter of the two stars.
These donor stars, which are in thermal equilibrium, can be significantly cooler than the single-star terminal age MS. 
In Fig.\,\ref{fig:HRD6}, we see these models sticking out on the cool side of the RBF about 0.5\,dex below the turnoff.
We note that such systems have been identified by \citet{2015A&A...582A..73H} and \citet{Mahy2020}.
Finally, our merged stars behave similar to near-conservative Case\,A mass-transfer systems. They populate the brightest part of the
extended turnoff region and overlap strongly with the post-mass-transfer mass gainers (Fig.\,\ref{fig:HRD6}).

The group of post-mass-transfer systems is dominated by initially wider binaries, which undergo thermal timescale
mass transfer only after the primary star exhausted hydrogen in its core (Case\,B mass transfer).
As their large orbital separations render tides ineffective on the mass gainers, 
these are quickly spun up to near-critical rotation, after which their accretion efficiencies are strongly reduced (see Section\,\ref{sec:method}).
Many Case\,B mass gainers end up naturally in this situation, as critical rotation is reached after only
accreting a few percent of their initial mass \citep{1981A&A...102...17P}, if the 
orbits are wide enough to avoid tidal spin-down \citep{2012ARA&A..50..107L}. 
These models populate the RBF, and are approximately 15\% cooler than the correspondingly luminous models on the SSI
due to the action of their centrifugal force on their structure \citep{2019ApJS..243...10P}. 
The RBF becomes more distinct and spreads over a larger luminosity range for larger ages. 
The RBF shows a temperature offset from the SSI of approximately 15\%. Our models are spun up to as fast a rotation
as is numerically allowed, i.e., 98\% of critical rotation, and they mostly remain at this level for their remaining
hydrogen burning lifetimes. Only the most massive fast rotators ($M \simgr 20\mso$) have strong enough stellar winds 
to spin down \citep{1998A&A...329..551L}. This leads them to slowly evolve off the RBF towards the blue side.
The RBF extends a factor of 30 or more in luminosity below the turnoff region. 
We show in Appendix \ref{sec:append2} that the minimum luminosity of the RBF is set by the 
minimum initial mass ratio for which a merging of the two stars is avoided. 

\begin{figure}
\begin{interactive}{animation}{animation.mp4}
\plotone{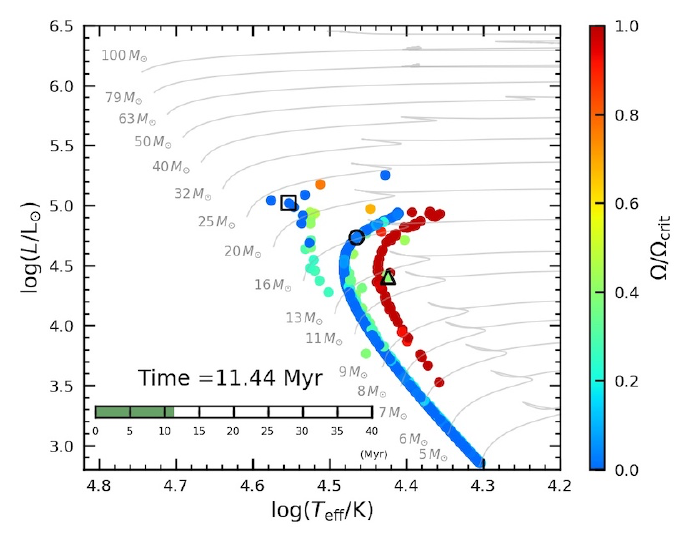}
\end{interactive}
\caption{Star cluster animation to show their evolution in the Hertzsprung-Russell diagram, with colors indicating
the ratio of rotation to critical rotation velocity. Evolutionary tracks of non-rotating single stars are shown as gray lines,
with initial masses quoted. We evolve our cluster from an age of 0.02\,Myr up to 40\,Myr, corresponding to a turnoff mass of $\sim 8\,\mathrm{M}_{\odot}$.
The three example binaries shown in Fig.\,\ref{fig:Ex} are marked with special symbols, such that triangles, circles and
squares indicate examples a, b, and c, respectively. The video duration is 20 s. The animation can be downloaded from the link for ancillary files or on the online journal \url{https://doi.org/10.3847/2041-8213/ab6171}.
\label{fig:video}}
\end{figure}

We provide an animation of our population synthesis study (Fig.\,\ref{fig:video}) to demonstrate the time evolution of 
the various binary-induced morphological MS features of young star cluster. 
In this animation, different colors correspond to different fractions of critical rotation. 
A slow forwarding of the image sequence therefore allows tracing of the evolution of the 
brighter stars in the HR diagram, as well as the evolution of their rotation velocities.

\section{Comparison to single-star models}
\begin{figure}
\centering
\includegraphics[width=0.7\linewidth]{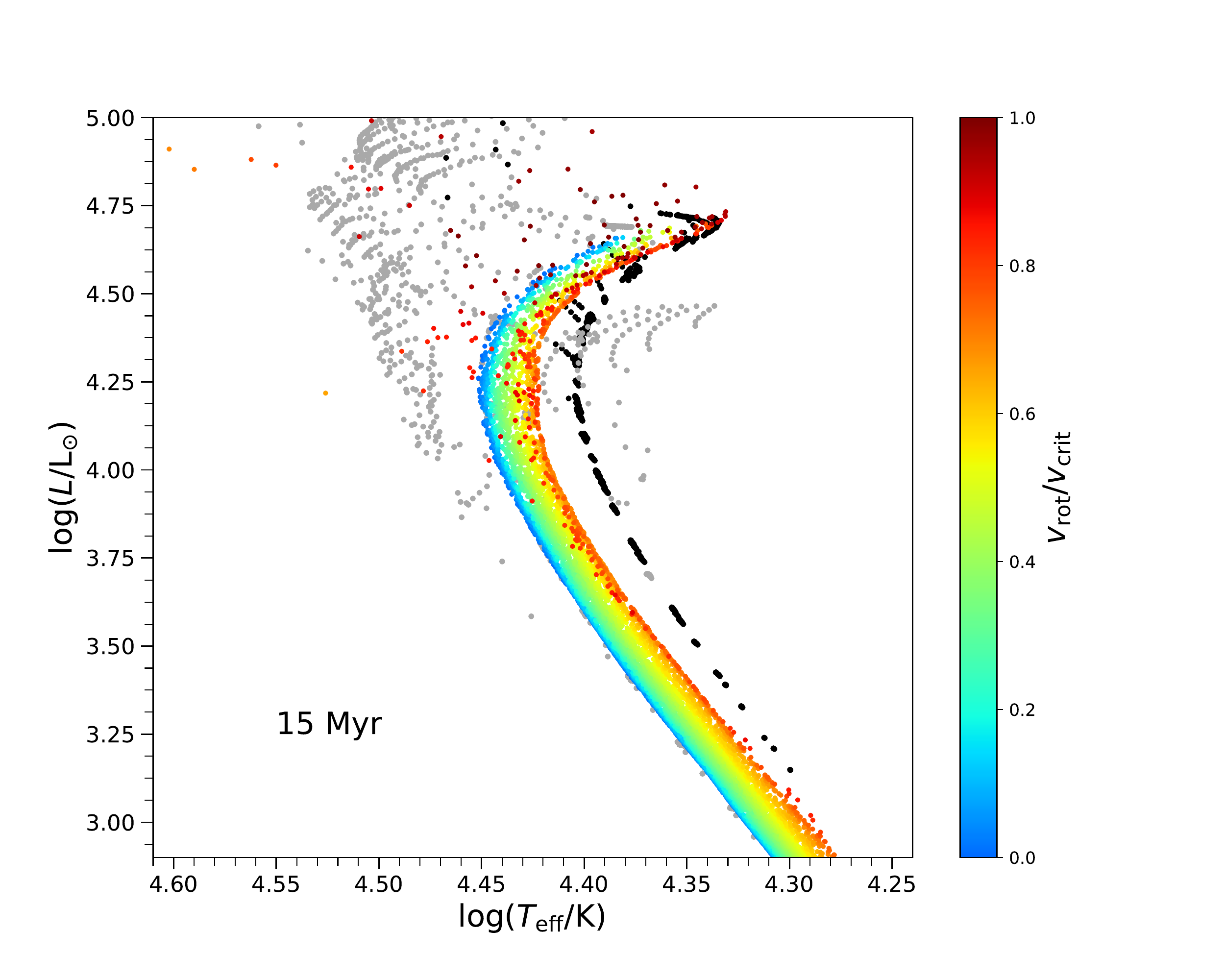}
\caption{Distribution of 10,000 simulated single stellar models (colored dots) obtained from interpolating in the dense grid of 
the single-star evolution model of \citet{2011A&A...530A.115B}, taking into account the initial mass function and the 
distribution of the initial rotation velocities derived by \cite{2013A&A...550A.109D}, at 15\,Myr. 
The color represents the ratio of rotation to critical rotation velocity at an age of 15\,Myr, according to the color bar to the right.
Overlaid are our binary-evolution models models using grey and black dots, at the same age, with black indicating a ratio of rotation 
to critical rotation velocity above 0.98. 
\label{fig:15myr_single_binary}}
\end{figure}

Figure\,\ref{fig:15myr_single_binary} shows the result of a population synthesis 
for single-star models \citep{2011A&A...530A.115B} with an age of 15\,Myr, for which we adopted the initial 
rotational velocity distribution of \cite{2013A&A...550A.109D} and the Salpeter IMF.
It demonstrates that rapidly rotating single-star models are significantly bluer than the spun-up mass gainers. 
Since the rotation rate of the RBF models was ordinary before accretion, 
they have established a strong chemical gradient in their interior, which prevents rotational mixing from affecting their chemical structure. 
The situation is different in rapidly rotating single stars, where helium mixed into the stellar envelopes \citep{2011A&A...530A.115B,2013A&A...558A.103G} can cause the star to evolve to a higher effective temperature \citep{1987A&A...178..159M}. 
Furthermore, rapidly rotating stars would only form a discrete MS line 
if their initial rotation rates were nearly identical, since the amount of mixing is a sensitive function of the rotation rate.

Figure\,\ref{fig:15myr_single_binary} shows also that rotational mixing in single stars can produce an extended turnoff
region similar to the TBF. However, the presence of the single-star feature depends on the number of stars that rotate
very rapidly initially. If the rapidly rotating single stars were in fact binary products, as suggested by \citet{2013ApJ...764..166D} and \citet{2015A&A...580A..92R}, binaries would be
the only way to produce an extended MS turnoff. As shown by our models, a TBF is in fact unavoidable
for a non-vanishing binary fraction. 

\section{Comparison to observed star clusters}
Within the last decade or so, it has become evident that extended MS turnoffs are a ubiquitous 
feature of young star clusters \citep[][and references therein]{2019ApJ...876...65L}. In a broader sense, this
includes the phenomenon of the blue stragglers, which is found for clusters of all ages \citep{Ahumada2007}. For young clusters, extended turnoffs are often explained by the presence of stars with
a wide range of rotation rates \citep[][and references therein]{2017NatAs...1E.186D}.  However, simplified
binary-evolution population synthesis demonstrates that binary evolution may also be able to lead to extended turnoffs \citep{2011ApJ...731L..37Y,2014ApJ...780..117S}.

\begin{figure}
\centering
\includegraphics[width=0.7\linewidth]{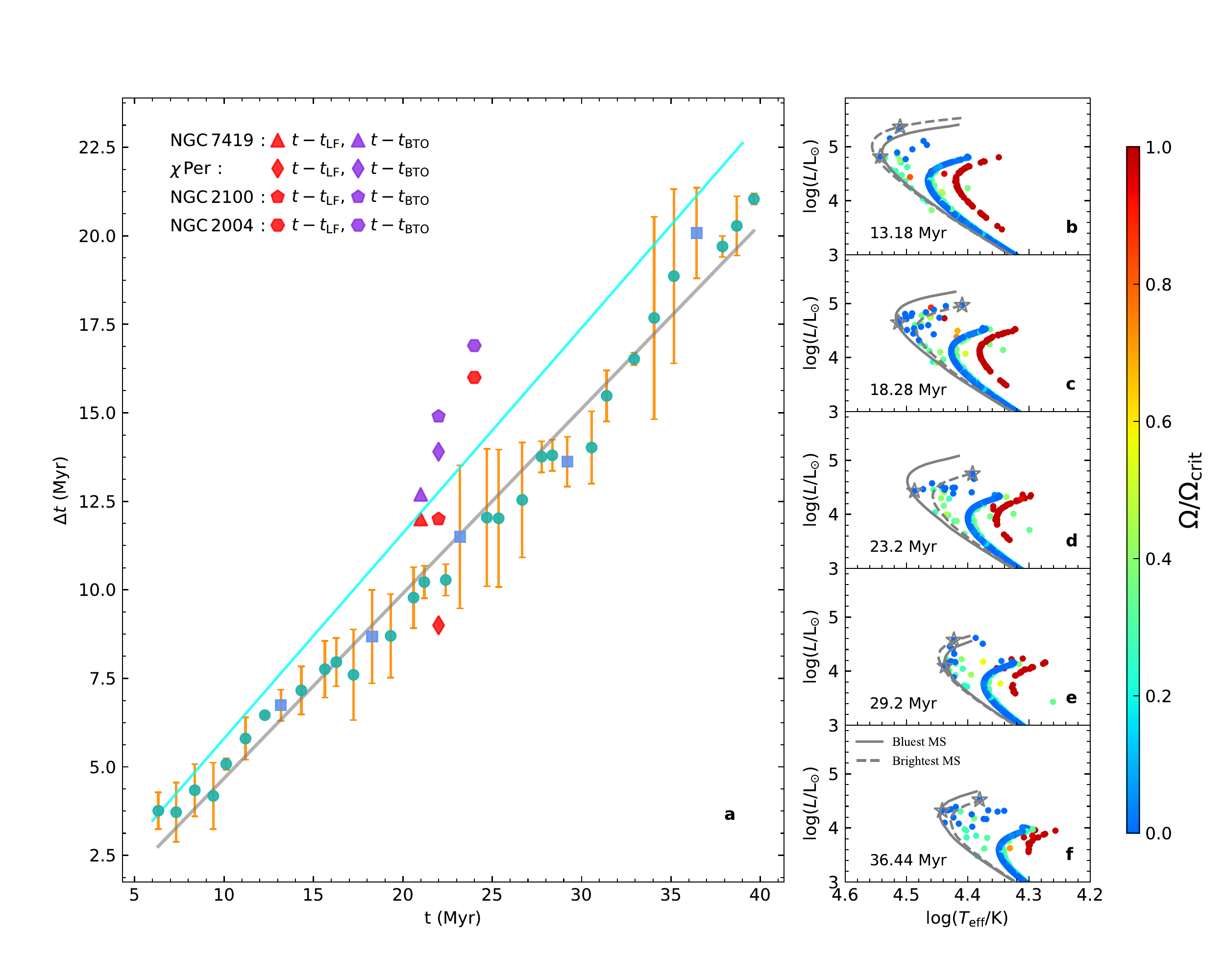}
\caption{
Age spread $\Delta\,t$ inferred by the extended MS turnoff from the binary models as a function of cluster age $t$.
To determine the apparent age of the extended turnoff stars in
our simulated clusters, we first generate a series of single-star isochrones and then fit these isochrones
via two methods, first, using the bluest MS star, and second, using the brightest MS star
as to be fitted by the isochrone. Examples are shown in Panels\,(b) to\,(f), where the solid gray line describes the best-fitting isochrone for the first, and the dashed gray line describes the best-fitting isochrone for the second method.
The model stars used for these fits are indicated by grey star symbols. In Panel\,(a) we show
average of the result from the two methods for 33 different times as green symbols, where the examples from Panels\,(b) to\,(f)
are marked by blue squares. The error bars represent the difference in the results from the two methods.
The gray straight line represents a linear fit to the data as $\Delta t = 0.518 t - 0.510$.
We also show the results for two Galactic and two LMC clusters of \citet{2019MNRAS.486..266B}.
Here, we took the age derived from the lowest luminosity red supergiant as the true cluster age
\citep{2019MNRAS.486..266B,2019A&A...624A.128B}, and show the age discrepancy with the ages derived
from the stellar luminosity function (red) and the brightest
turnoff star (purple). We display our analytic estimate (Eq.\,\ref{e:dt}) with a cyan line.
\label{fig:dt}}
\end{figure}

Our binary-evolution models allow us to obtain a reliable 
quantification of the extent of the TBF. We derived the apparent age difference
$\Delta t$ between the SSI and an SSI that fits either
the bluest or the brightest member of the coeval binary population in our simulated cluster (see Fig.\ref{fig:video}), 
as a function of its true age $t$. Figure\,\ref{fig:dt} shows that the apparent age difference can be approximated 
by a linear function, i.e., $\Delta t \propto  t/2$. This corresponds well to our analytic estimate in Appendix \ref{sec:append1}
of $\Delta t \propto 0.58 t$. As shown in Fig.\,\ref{fig:dt}, our result is compatible with the recent observational estimates
 by \citet{2019MNRAS.486..266B} and \citet{2019A&A...624A.128B}. Our analytic result also agrees with corresponding estimates
for clusters up to 1000\,Myr \citep[see Figure\,2 of][]{2015MNRAS.453.2070N}. 

In fact, any process that enriches the core with a fixed fraction of the hydrogen mass of the envelope
of a MS star will give rise to a linear relation between the apparent age difference at the turnoff
and the turnoff age of the non-enriched stars, be it binary rejuvenation as in our case, or rotationally induced mixing
\citep{2015MNRAS.453.2070N}.
However, our simple result, i.e., that the apparent cluster age is about half the true age,
is a natural consequence that binary evolution can lead to stars with a mass of about twice the single-star
turnoff mass. While rotational mixing can lead to a similar result, a tuning of the rotation rate
and/or of the mixing efficiency is required, since the amount of mixed material depends on both parameters.
In any case, it appears likely that binary evolution provides a significant, perhaps dominant, contribution 
to the extended turnoffs in young star clusters.

The occurrence of the RBF in our models is a genuine binary-evolution effect and does not depend
on the choice of initial rotation rates. 
The effectiveness of accretion-induced spin-up in nature is demonstrated by the observed group of Be/X-ray binaries, which are understood as
rapidly rotating B stars that have accreted from the progenitors of neutron star companions \citep{2018ApJ...853..156W}.
We therefore identify our RBF models, which all rotate extremely rapidly (see Fig.\,\ref{fig:video}), as Be stars
\citep{1991A&A...241..419P}.
Recent photometric and spectroscopic results show that a significant fraction of the MS stars
in young clusters ($\simle 100\,$Myr) within two to three magnitudes below the turnoff are Be stars
\citep{Milone2018,bodensteiner2019young}. As demonstrated by our animation (Fig.\,\ref{fig:video}),
noting that a difference of 2.5 mag corresponds to a factor of 10 in luminosity, 
this brightness range is well reproduced by our binary models. In Sect.\,\ref{sec:append2}
we relate this range analytically to the minimum mass ratio of stable Case\,B mass transfer,
which thus can be directly determined from the quoted observations.  

It has been shown by \citet{2008A&A...478..467E} that if stellar winds are not too strong,
rotating single stars increase their fraction of critical rotation during the MS evolution, which may
lead the initially fastest rotators to become Be stars. However, the transition to the Be stage occurs mostly at the end of core hydrogen burning. As shown by \citet{Hastings2020}, this would place the single Be stars very close to the cluster
turnoff, which seems to disagree with the quoted observations. The only way to obtain single Be stars 
a factor of 10-30 below the turnoff luminosity in star clusters is that they are born with
near-critical rotation. According to \citet{2005ApJS..161..118M}, the number
of such stars appears to be small. We conclude that it appears likely that binary evolution provides 
a significant, perhaps dominant, contribution to the Be star population in young star clusters, as suggested by \citet{2018A&A...615A..30S} and \citet{2019ApJ...885..147K}. 

%
%
\section{Concluding remarks}
Our binary-evolution models produce several distinct morphological features of the MSs of star 
clusters younger than about 40\,Myr. 
Given that our results emerge from generic binary-evolution model calculations, without
applying any fine-tuning, it is striking that they quantitatively reproduce several of the main characteristics of observed 
MSs of open star clusters. This refers in particular to the extended turnoff region and
the corresponding apparent age spread, as well as to the luminosity range of the Be stars in young
clusters. Both effects are comprehensively demonstrated by our animation (Fig.\,\ref{fig:video}).


This does not preclude the possibility that other factors are also important. As discussed above,
the role of rapidly rotating single stars depends critically on the initial distribution of rotation rates.
Also, non-coevality or differences in the initial chemical composition of the cluster stars may be important in some clusters. 
Our results imply that binary evolution is likely to play a major role in shaping the MSs of star clusters,
as long a substantial fraction of the stars are born in binary systems.
A detailed population synthesis study including all of the above factors will help us to confirm this picture, 
and will have strong implications for our understanding of star and binary formation and evolution,
including the origin of the mass ratio and orbital period distribution of close binaries, and the IMF of stars. 

\acknowledgments
C.W. is thankful for the financial support from the CSC scholarship. P.M. acknowledges support from NSF grant AST-1517753 
and the Senior Fellow of the Canadian Institute for Advanced Research (CIFAR) program in Gravity and Extreme Universe, 
both granted to Vassiliki Kalogera at Northwestern University.

\appendix
\section{Appendix: Quantitative characterization of the main-sequence features }\label{sec:Appendix}
We quantify the features of the TBF, RBF, and BBF described in Fig. 1 in Appendix \ref{sec:append1}, \ref{sec:append2}, and \ref{sec:append3}, respectively.
\subsection{TBF (Turnoff binary-evolution feature)}\label{sec:append1}
 The apparent age spread indicated by the extended turnoff region is about half of the true age of the simulated stars 
\citep[see Figure 2 of ][]{2015MNRAS.453.2070N}. This compares well with simple estimates 
of the binary rejuvenation due to mass accretion and stellar mergers. 
For a mass-luminosity relation as $L \propto M^{\alpha}$, and a hydrogen burning lifetime of a star of mass $M$ as
$\tau \propto M / L$, the lifetime ratio of two stars is
$\tau_{\rm 1}/\tau_{\rm 2}=(M_{\rm 1}/M_{\rm 2} )^{1-\alpha}$, while their luminosity ratio is
$L_{\rm 1}/L_{\rm 2}=(\tau_{\rm 1}/\tau_{\rm 2} )^{\alpha / (1-\alpha)}$.
The maximum possible stellar mass in a cluster with a single-star turnoff mass of 8$\mso$ 
(corresponding to $M_{\rm 1}$ in the equations above),
will be about 14.4$\mso$ ($= M_{\rm 2}$), i.e., the result a merger of two 8$\mso$ stars (90\% of $16\mso$).
Therefore, we obtain an equation for the apparent age spread $\tau_{\rm 1} - \tau_{\rm 2}$ in our cluster with an age $\tau_{\rm 1}$
 that is linear in $\tau_{\rm 1}$, as
\begin{linenomath*}
    \begin{equation}
    \tau_{\rm 1} - \tau_{\rm 2} = \left[ 1 - \left( M_{\rm 1} \over M_{\rm 2} \right)^{\alpha -1} \right]  \tau_{\rm 1},
    \label{e:dt}
    \end{equation}
\end{linenomath*}

and for $M_{\rm 1}/M_{\rm 2}=8/14.4$ and for $\alpha = 2.5$, which holds approximately for stars 
of 14$\mso$ and is only a weak function of the mass \citep[see Figure \,17 of][]{2015A&A...573A..71K},
we obtain $\tau_{\rm 1} - \tau_{\rm 2} = 0.58 \tau_{\rm 1} $.


\subsection{RBF (Red binary-evolution feature)}\label{sec:append2}
We note that our MESA models may underestimate the effect of the centrifugal force on the 
outermost parts of the stellar models   
for a rotation faster than about 90\% of critical rotation \citep{2019ApJS..243...10P}.         
Furthermore, they do not account for the presence of a decretion disk, which may make the star appear
redder than predicted here \citep{2013A&ARv..21...69R}. This does induce an uncertainty in the
predicted effective temperature, or any magnitude or color computed from it.

The extent of the RBF in luminosity is determined by the minimum mass ratio for stable mass transfer. 
For example, at a turnoff masses of 20$\mso$ (corresponding to $t\simeq$10\,Myr),
the minimum initial mass ratio $q_{\rm min,B}$ for which a merger during the Case\,B mass transfer is avoided  
is $q_{\rm min,B}\simeq 0.3$, i.e., the minimum stellar mass on the RBF is about $6\mso$
(as little mass is accreted). Just from the mass difference, using the mass-luminosity relation
of the form $L \propto M^{\alpha}$ with $\alpha\simeq 2.5$ \citep{2015A&A...573A..71K}
implies a luminosity ratio between the brightest and the dimmest RBF star of 20 (1.31 dex). 
However, due to the large mass difference, while the brightest RBF star is about to finish core hydrogen
burning, the dimmest ones are essentially unevolved. We therefore need to consider the difference in
the average mean molecular weight $\mu$ of the stars, which contributes to the luminosity difference
as $L \propto \mu^{\beta}$. An initial mean molecular weight of $\mu_{\rm SMC}=0.59$, the
mean molecular weight of helium as $\mu_{\rm He}=4/3$, and an average convective core mass fraction of
$q_{\mathrm{core}}=0.25$, yield an average mean molecular weight of our brightest RBF star of $\bar \mu = 0.69$,
which, with $\beta \simeq 4.5$ \citep[see Figure\,17 of ][]{2015A&A...573A..71K}, yields
another factor of 2.02 (0.31 dex) for the luminosity ratio, resulting in a total of 41 (1.61 dex),
i.e., in more general terms, the luminosity ratio between the brightest and the dimmest RBF star is
\begin{linenomath*}
    \begin{equation}
    {L_{\rm top} \over L_{\rm bottom}} =  \left({\mu_{\rm He} \over {{\mu_{\rm SMC}} q_{\rm core}
    + \mu_{\rm He} (1-q_{\rm core})}} \right)^{\beta}  \left( 1\over q_{\rm min,B} \right)^{\alpha}.
    \label{e:qmin}
    \end{equation}
\end{linenomath*}
Considering the dependence of $q_{\rm min,B}$ on cluster age, i.e., 
for $t \simle 15\,$Myr we have $q_{\rm min,B}\simle 0.3$, which rises to $q_{\rm min,B}\simle 0.6$
at 20\,Myr and 0.7 at 30 Myr and beyond, the above equation leads to a good characterization of the RBF
for the different times shown in Fig.\,\ref{fig:HRD6}.

Due to mass accretion during Case\,B, the fraction of RBF stars near the
turnoff region will be naturally close to one, as observed in many young clusters.
At lower luminosities, the RBF star fraction will be proportional to the fraction of Case\,B
binaries that avoid merging.

The temperature separation between RBF and SSI cannot be predicted well by our models.
The reason is that the RBF stars rotate at more than 0.98\% of critical
rotation. The outer layers of current rotating 1D models are not well described for models beyond
0.90\% of critical rotation \citep{2019ApJS..243...10P}. 

\subsection{BBF (Blue binary-evolution feature)}\label{sec:append3}
The BBF merges with extended turnoff stars at the highest luminosities. The minimum luminosity
is determined in our models by the minimum initial mass ratio for which Case\,A contact evolution is avoided for
at least a fraction of core hydrogen burning, which is  
$q_{\rm min, A}=0.7$ for an initial donor mass of $M_{\rm 1,i}=10\mso$, and $q_{\rm min, A}=0.35$ 
for an initial donor mass of $M_{\rm 1,i}=20\mso$. 
We note that our predictions for the evolution of contact systems must be considered as very uncertain,
as our models do not allow for heat flows between the two components. 
Therefore, the more realistic values of $q_{\rm min, A}$, as well as the number of blue MS stars, may
differ significantly from those obtained in our calculation.





\end{document}